\documentclass{desyproc}

\def\be{\begin{equation}}
\def\ee{\end{equation}}
\def\bea{\begin{eqnarray}}
\def\eea{\end{eqnarray}}

\def\Mpl{M_\mathrm{Pl}}
\def\Ve{V_\mathrm{eff}}

\def\be{\beta_\mathrm{eff}}
\def\bmat{\beta_\mathrm{m}}
\def\bg{\beta_\gamma}
\def\gsim{ \lower .75ex \hbox{$\sim$} \llap{\raise .27ex
\hbox{$>$}} }
\def\lsim{ \lower .75ex \hbox{$\sim$} \llap{\raise .27ex
\hbox{$<$}} }

\begin{document}
\title{Hunting for Chameleons in ALP searches}

\author{{\slshape Amanda Weltman$^{1,2}$}\\[1ex]
$^1$Department of Applied Mathematics and Theoretical Physics, Cambridge University,\\Cambridge CB2 0WA, United Kingdom.\\ 
$^2$Department of Mathematics and Applied Mathematics,\\ University of Cape Town, Private Bag, Rondebosch, South Africa, 7700}

\maketitle

\begin{abstract}
We discuss some recent developments in chameleon models presented at the Fourth Patras workshop on Axions, WIMPs and WISPs. In particular we discuss the possibility of searching for chameleons in axion-like particle searches performed in the laboratory. Such chameleons may couple to both photons and matter with different coupling strengths. We discuss the exciting possibility of searching for these dark energy candidates in quantum vacuum experiments - in particular for the GammeV experiment at Fermilab.\end{abstract}

\section{Introduction}

Without doubt, one of the most riveting problems of modern physics is the cosmological constant problem. The remarkable observation that the universe is accelerating in its expansion has pushed theorists to propose ever more creative theories while at the same time pushing observational cosmologists to probe the detailed nature of so-called Dark Energy with increasingly sophisticated equipment and techniques.\\

\noindent
In this short article, based on a talk given at the $IV^{th}$ Patras Workshop, we will discuss one of the many Dark Energy models that is perhaps most compelling because of the predictions it makes for non-cosmological experiments. So-called chameleon models \cite{chamKW} can be tested both in space and in the laboratory in varied instantiations. These complementary probes allow us to learn about a dark energy model without any cosmological measurements within the settings of experiments that are all designed for other purposes. As such they provide both added motivation and a low cost per physics output quotient for such experiments. 
 
\section{Chameleon Theories}

Chameleon fields have been introduced in recent years \cite{chamKW,chamcos} to allow for cosmologically evolving light scalar fields that do not violate any known local fifth force or Equivalence Principle tests. Most enticingly, they offer a dark energy candidate whose properties may be probed in our local environment; via local tests of gravity in space and tests of quantum field theory in the lab. The essential feature of these fields is that due to their couplings to matter they acquire an effective potential and thus an effective mass that depends quite sensitively on the background energy density. In this way they have managed to satisfy all tests of the equivalence principle and fifth force experiments so far \cite{chamKW,Upadhye}. Not only can these fields use their local environment to hide from our tests but their properties change depending on their environment thus allowing for both nontrivial predictions for tests of gravity in space and for cosmological effects. For extensive details on chameleon physics see \cite{chamKW,chamcos,chamstrong}.

\subsection{Ingredients}

In \cite{chamcos} an action of the following form is proposed for Chameleon fields coupled to matter and photons; 
 \begin{equation}
S = \int d^4x \sqrt{-g} \left(\frac{1}{2M_{pl}^2} R - \partial_\mu \phi \partial^\mu \phi - V(\phi)\right) 
  - \frac{e^{\phi/M_{\gamma}}}{4}F^{\mu\nu}F_{\mu\nu} + S_m(e^{2\phi/M^i_m} g_{\mu \nu}, \psi^i_m) \label{e:action} \,\, ,
\end{equation}
where $S_m$ is the action for matter and in general the chameleon field, $\phi$ can couple differently to different matter types $\psi_i$, and $V(\phi)$ is the chameleon self interaction.  For simplicity here we will consider a universal coupling to matter defined by $\beta_m = \Mpl/M_m$ while allowing for a different coupling to electromagnetism, $\beta_{\gamma} = \Mpl/M_{\gamma}$, through the electromagnetic field strength tensor $F_{\mu \nu}$. \\

\noindent
The non-trivial coupling to matter and the electromagnetic field induces an effective potential 
\begin{equation}
\Ve(\phi,\vec x) = V(\phi) + e^{\beta_m\phi/\Mpl} \rho_m(\vec x) + e^{\beta_\gamma\phi/\Mpl} \rho_\gamma(\vec x),
\end{equation}
where we have defined the effective electromagnetic field density $\rho_\gamma = \frac{1}{2}(|\vec B^2-|\vec E|^2)$ rather than the energy density. An essential insight of chameleon models is noticing that the presence of matter and electromagnetic fields induces a minimum $\phi_\mathrm{min}$ in $\Ve$ where $V$ can be a monotonic function. The dependence of this minimum on the background matter and electromagnetic fields causes the effective mass of the chameleon field to change in response to its environment.  In turn we find varied chameleon phenomenology depending on the experimental setup and hence the environment. \\

\noindent
We can see explicitly that for an exponential potential, the effective mass of the field $\phi$ is dependent on the local density of matter and electromagnetic fields,
\begin{equation}
V(\phi) = \Lambda^4 \exp\left(\frac{\Lambda^n}{\phi^n}\right) \label{exp},    \quad 
\phi_\mathrm{min} \approx \left(\frac{n \Mpl \Lambda^{n+4}}{\bmat \rho_m + \bg \rho_g}\right)^{\frac{1}{n+1}} \, \, 
{\rm and} 
\,\,\, m_\phi^2 \approx  \frac{(n+1)} {\left(n \Lambda^{n+4}\right)^{\frac{1}{n+1}}} \left(\beta_m \rho_m +  \bg \rho_\gamma \right)^{\frac{n+2}{n+1}}   \\ \nonumber
\end{equation}
where the next to leading order terms are suppressed by factors of $\beta_i \phi / \Mpl  \ll 1$.

\section{Tests in the Lab : The GammeV Experiment}

In their initial realisation \cite{chamKW}, chameleon theories held the enticing promise of being observable in space based tests of gravity. Such models studied scalar fields coupled with gravitational strength to matter, i.e. $\beta_m \sim \mathcal{O}(1)$. Their unique properties explained why they have not yet been observed in earth-based tests of gravity. However, in recent years, it has been realised that even very strongly coupled chameleons, i.e. $\beta \sim \mathcal{O} (10^{10})$ even could evade tests of gravity on earth \cite{chamstrong}. While strongly coupled models are essentially not testable in space they do open up the possibility for tests in the laboratory to reveal chameleons. Quantum vacuum experiments hold the promise for observing chameleons or constraining strongly coupled chameleons. In particular searches for axions or axion like particles (ALPS) through photon to ALP conversion in the presence of a magnetic field offer a window  into constraining strongly coupled chameleon theories \cite{gammevwebpage, chamgammev, jar, alpenglow, gies}. A word of caution - it is important to realise that even though such experiments are probing the chameleon-photon coupling, $\beta_\gamma$ (and not the matter coupling directly), at their current stage, they can only constrain a theory that has strong coupling to both photons and matter \cite{gammevcham}. Future experiments may broaden this range to include $\mathcal{O}(1)$ couplings to matter, which would be most interesting \\

\noindent
Consider a vacuum chamber, whose walls, of density $\rho_\mathrm{wall}$, are much thicker than the chameleon Compton radius associated with the density $\rho_\mathrm{wall}$. Chameleons in the vacuum will be nearly massless and unaffected by the chameleon field in the chamber wall. As a chameleon particle approaches the wall, its mass will increase.  If its momentum is less than the chameleon mass inside the wall, then it will bounce elastically off the wall.  Thus the vacuum chamber will serve as a ``bottle'' for chameleon particles.\\

\begin{figure}[hb]
\centerline{\includegraphics[width=0.65\textwidth]{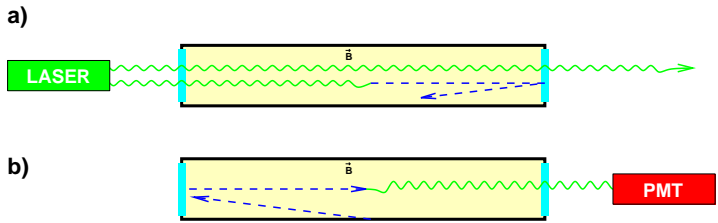}}
\caption{Schematic of the GammeV apparatus. a) Chameleon production phase: photons propagating through a region of magnetic field oscillate into chameleons. Photons travel through the glass endcaps whereas chameleons see the glass as a wall and are trapped. b) Afterglow phase: chameleons in the chamber gradually decay back into photons and are detected by a photomultiplier tube.}
\label{Apparatus}
\end{figure}

\noindent
The GammeV experiment at Fermilab \cite{gammevwebpage} is just such a chameleon bottle \cite{chamgammev}.  If the chameleon field couples to photons as well as to baryonic matter, then a photon interacting with a magnetic field in the vacuum chamber will oscillate into a chameleon particle, analogous to an ALP.  When this superposition of photon and chameleon states hits a glass window on one side of the chameleon bottle, it will be measured in the quantum mechanical sense; photons pass through the glass window, while chameleons are reflected.  A continuous source of photons entering the bottle will gradually fill it with a gas of chameleon particles.  After the photon source is turned off, chameleons will decay back into photons, which can escape the bottle through the glass window.  GammeV looks for this ``afterglow'' effect, a unique signature of photon-coupled chameleon particles \cite{jar}. See figure \ref{Apparatus} for a schematic of the experiment apparatus. For a more detailed discussion of the GammeV experiment and the first results see \cite{gammevwebpage, gammevcham}.

\section{Conclusions}
  
In this short article we have reviewed the intriguing possibility that a scalar field may couple directly to both matter and photons. The rich phenomenology of these chameleon models makes concrete predictions for tests in space as well as in the laboratory possible providing us with complementary ways of testing chameleon theory by probing different regions of parameter space. Most excitingly, there is a very real possibility that within the coming decade chameleon fields could either be observed or ruled out entirely using the space tests, laboratory tests and astrophysical observations becoming available to us. It is worth emphasising that regions of parameter space that have already been ruled out for axions will not necessarily be ruled out for chameleons. This was the lesson of \cite{chamPVLAS} where it was shown that the CAST experiment and the original PVLAS signal are perfectly compatible for chameleons though not for axions. The nature of chameleon fields - i.e. their differing mass depending on environment means that we cannot blindly apply the regions of exclusion from axion searches but rather these studies need to be readdressed to apply constraints to the chameleon parameter space.

\section*{Acknowledgments}
A.W would like to thank the Organisers and the participants of the Fourth Patras Workshop for truly excellent and productive meeting. A.W. gratefully acknowledges her many collaborators on Chameleon models in particular - Justin Khoury, Anne Davis, Phillipe Brax, Carsten van de Bruck and most recently the GammeV collaboration, especially Amol Upadhye and Aaron Chou for many useful discussions and for their insights onto chameleon phenomenology. 

\begin{footnotesize}

\end{footnotesize}


\begin{thebibliography}{99}
\bibitem{jar} The ``particle trapped in a jar'' technique was developed as part of the GammeV experiment and independently realised in \cite{alpenglow, gies}
\end{thebibliography}
\end{document}